\documentclass[aps,prd,eqsecnum,showpacs]{revtex4}
\usepackage{epsfig}

\begin{document}

\preprint{}

\title{Tuning gravitational-wave detector networks to measure compact
binary mergers}
\author{Scott A.\ Hughes}
\email{hughes@kitp.ucsb.edu}
\affiliation{Kavli Institute for Theoretical Physics, University of
California, Santa Barbara, CA 93106}

\date{\today}

\begin{abstract}
Gravitational waves generated by the final merger of compact binary
systems depend on the structure of the binary's members.  If the
binary contains neutron stars, measuring such waves can teach us about
the properties of matter at extreme densities.  Unfortunately, these
waves are typically at high frequency where the sensitivity of
broad-band detectors is not good.  Learning about dense matter from
these waves will require networks of broad-band detectors combined
with narrow-band detectors that have good sensitivity at high
frequencies.  This paper presents an algorithm by which a network can
be ``tuned'', in accordance with the best available information, in
order to most effectively measure merger waves.  The algorithm is
presented in the context of a toy model that captures the qualitative
features of narrow-band detectors and of certain binary neutron star
merger wave models.  By using what is learned from a sequence of
merger measurements, the network can be gradually tuned in order to
accurately measure the waves.  The number of measurements needed to
reach this stage depends upon the waves' signal strength, the number
of narrow-band detectors available for the measurement, and the
detailed characteristics of the waves that carry the merger
information.  Future studies will go beyond this toy model,
encompassing a more realistic description of both the detectors and
the gravitational waves.
\end{abstract}

\maketitle

\section{Introduction}
\label{sec:intro}

Much of the promise of gravitational-wave (GW) observation is in its
potential as a novel probe of physics and astrophysics.  Because GWs
couple very weakly to matter and arise solely from gravitational
interactions, sources that are completely dark electromagnetically may
be strong GW emitters.  By tracking and measuring the waves generated
in violent astrophysical events, we may gain insight into physical
processes that cannot be easily measured in other ways.

One example of such hard-to-measure physics is the late merger of
compact binary systems.  GW emission carries energy and angular
momentum out of the binary, driving the compact bodies ever closer
together.  During the early {\it inspiral} portion of this coalescence
process, the structure of the binary's members plays little role; they
can be usefully approximated as point masses, or spinning point
masses.  As the bodies come closer together, their internal structure
becomes very important.  The GWs generated in the final stages of
inspiral and {\it merger}, when the bodies collide and merge into some
coalesced state, will carry information about the bodies' structure.

It has long been recognized that, if the binary contains at least one
neutron star, merger waves will depend on the nature of neutron star
matter (cf.\ Refs.\ {\cite{clark_eardley,300yrs,3minutes,vallis}} and
references therein).  This opens the exciting possibility that GW
measurements by detectors such as LIGO {\cite{ligo}} could study the
properties of very dense matter, such as its equation of state (EOS)
{\cite{3minutes,klt,lee_eos}}.  Of particular interest will be testing
whether neutron stars contain a core of ``exotic'' matter, such as a
free quark state of some kind.  Models of compact stars comprised of a
free quark fluid, or with a free quark core, can have a structure
rather different from ordinary neutron stars --- they are often of
smaller radius, and there may exist a sharp density transition at some
finite radius from the star's core {\cite{alford,arrw}}.  If one or
both members of a binary system had such a structure, there could be
an observable imprint on the merger GW signature.  (We should
emphasize, though, that much work remains to evaluate whether such a
signature exists, and if so, what is its nature.)

Simulations of binary neutron star merger using Newtonian gravity and
a polytropic EOS (see, e.g., {\cite{zcm,rs}}) have found that the
merger waves indeed carry information about the EOS, but that these
waves are at very high frequency ($f\sim 1500 - 3000\,{\rm Hz}$) where
broad-band LIGO-type detectors do not have good sensitivity.  More
recent work using irrotational matter configurations in the conformal
approximation to GR {\cite{fgrt}} shows that the EOS-dependent
structure is likely to come out at somewhat lower frequencies, $f\sim
1000\,{\rm Hz}$; this is still high enough that broad-band detector
sensitivity is not very good.  If one member of the binary is a black
hole, the EOS-dependent information will come out at still lower
frequencies, $f\sim 400 - 1000\,{\rm Hz}$ {\cite{vallis}} --- the
larger system mass shifts all frequencies downward.  In this case,
broad-band detectors are more useful for studying the merger waves,
but still may not be ideal.

To learn as much as possible from the waves generated during the
merger, broad-band GW detectors should be supplemented by {\it
narrow-band} detectors.  Acoustic detectors in existence today
{\cite{blair,cerdonio}} and special interferometer topologies under
development such as signal recycling {\cite{sr}} and resonant sideband
extraction {\cite{rse}} have good sensitivity in a narrow band at high
frequencies, $\delta f/f\sim 0.1 - 0.2$ for $f\sim 500 - 2500\,{\rm
Hz}$.  Narrow-band detectors answer essentially a yes-no question:
``Did the binary radiate in my frequency band?''  A ``xylophone'' of
narrow-banded detectors would probe gross features of the merger
waveform, such as a sharp cutoff in the GW spectrum (seen in recent
simulations {\cite{fgrt}}), or the waves generated by a transient bar
that forms in the merger detritus (seen in some Newtonian simulations
{\cite{zcm,rs}}).  Such simple measurements should be robust in the
sense that they wouldn't require detailed modeling of the waves'
phasing --- very important, since theoretical uncertainites in the
merger waveforms are likely to be significant even when these
measurements can be made.

Practical considerations such as cost and available facility space
will limit the number of narrow-band detectors that can be used for
each measurement.  To make best use of these detectors, the network
should be designed in a way that is in some sense optimal: the
narrow-band detectors should be configured, in concert with the
broad-band detectors, so that the network of all detectors is most
likely to provide new information about merger waves, given our best
present knowledge of the waves' properties.  How one ``tunes'' a
detector network in this manner is the subject of this paper.

We assume that an inspiral has already been measured, so that we know
merger waves must be present in the data.  For the purposes of this
analysis, we assume further that the waveform depends on a single
parameter $\lambda$ that grossly characterizes the merger waves.  This
$\lambda$ could be the frequency of a sharp cutoff in the wave
spectrum, or the frequency at which a short-lived bar may radiate for
several cycles.  Theoretical modeling allows us to phenomenologically
relate this parameter to a description of the binary's stars.  For
example, in models in which the wave spectrum sharply cuts off,
$\lambda$ is most strongly related to the compactness of the stars:
smaller stars exhibit a cutoff at higher frequencies {\cite{fgrt}}
since they spiral in further before the spectrum cuts off.  Describing
the merger features with a single parameter is no doubt an
oversimplification, but is useful for demonstrating how network tuning
works to zoom-in on gross features of the waves.  Tuning the network
means finding the configuration which measures $\lambda$ with as
little error as possible.

We develop a tuning algorithm that does just this: it configures the
network to measure $\lambda$ as accurately as possible, given our
uncertainty in $\lambda$'s value, and updates (``retunes'') the
network as measurements teach us about merger waves.  This algorithm
is based on the maximum likelihood GW measurement formalism developed
by Finn {\cite{finn92}}.  Finn defines two probability distributions
which play a major role here: the {\it prior} probability,
$p_0(\lambda)$, summarizing all that is known about $\lambda$ before
measurement; and the {\it posterior} probability, $P_{\rm
post}(\lambda |\hat\lambda)$, summarizing what is known afterward.
The posterior distribution is built from the measured datastream, and
so explicitly depends upon the detector network's characteristics and
on the measured waveform $h(\hat\lambda)$, where $\hat\lambda$ is the
unknown, true value of $\lambda$.

Following measurement, the posterior probability is the tool one uses
to estimate the values of the parameters describing GWs, and also to
estimate the error in those values {\cite{finn92,finnchernoff,cf}}.
It is not quite the tool needed here: we need to estimate the accuracy
with which $\lambda$ is likely to measured, but we need this estimate
{\it before} measurement.  To this end, we introduce an additional
probability, the {\it anticipated} distribution of $\lambda$.  This is
a marginal distribution found by integrating out the dependence of the
posterior probability upon the unknown true parameterization
$\hat\lambda$:
\begin{equation}
P_{\rm ant}(\lambda) = \int p_0(\hat\lambda)P_{\rm post}(\lambda
|\hat\lambda)\,d\hat\lambda\;.
\end{equation}
This is the distribution that we we anticipate will describe $\lambda$
after a measurement.  Like the posterior probability, it depends on
the detector network, so we can tune the network's adjustable
parameters to find the network which we anticipate will measure
$\lambda$ as accurately as possible, given our current ignorance of
the merger waves.

From the anticipated distribution's definition, it is simple to update
and improve the network as we learn more about $\lambda$.  Following a
measurement, we construct the posterior distribution $P_{\rm
post}(\lambda |\hat\lambda)$ from the datastream.  We then use this
posterior distribution as the prior distribution for the next
measurement: $p_{0}^i(\lambda) = P_{\rm post}^{i-1} (\lambda
|\hat\lambda)$.  We update the anticipated distribution with the new
priors, and then update the network.  The detector network is thereby
adjusted and improved following each measurement, so that our ability
to measure $\lambda$ is incrementally improved by each measurement.
Our gradually improving knowledge of $\lambda$'s value is manifested
as a gradual peaking of this distribution: we begin with a prior
describing complete ignorance (uniform distribution) and find that
after several merger measurements the distribution begins to peak
around the true value $\hat\lambda$.  The more we learn about the
merger waves, the more sharply peaked becomes this distribution.

In this paper, we demonstrate the concept and principles of network
tuning using a toy model for the merger and for the narrow-band
detectors.  In Sec.\ {\ref{sec:formal}}, we describe in more detail
the probability distributions introduced above, and then discuss how
we use them to tune our network designs.  Our tuning procedure is
given explicitly at the end of this section.  We next present our toy
model in Sec.\ {\ref{sec:toy}}: we approximate narrow-band detectors
as zero bandwidth (delta function) GW detectors with adjustable center
frequency, and treat the waveform $h(t;\lambda)$ as a quadrupole
inspiral chirp up to a merger frequency $f_m\equiv\lambda\times
1000\,{\rm Hz}$.  This highly simplified description throws away
important physics, particularly the finite bandwidth expected in real
narrow-band detectors and the slower frequency rolloff seen in recent
merger computations (e.g., Ref.\ {\cite{fgrt}}).  It is simple enough,
though, that many of the calculations needed can be done analytically,
and is close enough to the real problem that it should give a good
sense of how network tuning is likely to proceed in practice.

We test our tuning algorithm in Sec.\ {\ref{sec:results}}.  We show
that measurements can converge on an accurate value for $\lambda$
after measuring some number of binary merger events.  The size of that
number depends on the signal strength (a few strong signals can drive
convergence rather quickly) and the number of narrow-band detectors
available for the measurement (having at least two available can speed
up convergence quite a bit).  The good behavior of our tuning
procedure can be taken as an indication that this algorithm is robust.
However, we cannot pretend that our analysis is in any way definitive:
the toy description of the detectors and the waveform neglects several
important effects.  This analysis should therefore be regarded as a
proof-of-concept presentation; future work, discussed in Sec.\
{\ref{sec:future}}, will put the various complications neglected here
back where they belong.  In particular, we plan to use more realistic
descriptions of the narrow-band detectors (cf.\ Refs.\
{\cite{whitepaper,hhs}}), and merger waveforms taken from recent
computational models.

\section{Formalism}
\label{sec:formal}

\subsection{Probability distributions}
\label{subsec:dist}

As mentioned above, we assume that we have already detected a GW via
the inspiral, and so we have already learned many interesting
parameters characterizing the binary --- combinations of the masses
and spins, the distance to the binary to some accuracy, etc.\
{\cite{cf,pw}}.  We now wish to measure the merger parameter
$\lambda$.

Consider a network of $N$ GW detectors.  We use the notation of
Appendix A of Ref.\ {\cite{bbhII}}: the datastream of the $j$-th
detector in the network is $g_j$; the output of the whole network is
${\vec g}$.  Because we assume that a GW has been detected, this
datastream consists of a GW signal ${\vec h}(\hat\lambda)$ plus noise
${\vec n}$: ${\vec g} = {\vec h}(\hat\lambda) + {\vec n}$.  (Recall
that $\hat\lambda$ represents the true, unknown value of $\lambda$
that describes the waves in the datastream.)  Our goal is to determine
the probability that, given this datastream, we will measure the
parameter $\lambda$.

Our first step is to find the posterior probability that we measure
the waveform ${\vec h}(\lambda)$ given the datastream $\vec g$.  Finn
has developed this probability for a single detector {\cite{finn92}};
using the notation of Ref.\ {\cite{bbhII}, the generalization to
multiple detectors is straightforward.  Following Finn, we begin with
Bayes' law:
\begin{equation}
P[{\vec h}(\lambda)|{\vec g}] = {P[{\vec g}|{\vec h}(\lambda)] P[{\vec
h}(\lambda)]\over P({\vec g})}\;.
\end{equation}
The probabilities written here have the same meaning as those used in
Eq.\ (2.3) of \cite{finn92}:
\begin{eqnarray}
P[{\vec h}(\lambda) | {\vec g}] &=& \mbox{the conditional probability
that ${\vec h}(\lambda)$ is present given the datastream ${\vec g}$}\;;
\nonumber\\
P[{\vec g} | {\vec h}(\lambda)] &=& \mbox{the conditional probability
of measuring the datastream ${\vec g}$ given ${\vec h}(\lambda)$}\;;
\nonumber\\
P[{\vec h}(\lambda)] &=& \mbox{the prior probability that the waveform
${\vec h}(\lambda)$ is present}
\nonumber\\
&\equiv& p_0(\lambda)\;;
\nonumber\\
P({\vec g}) &=& \mbox{the probability that the datastream ${\vec g}$
is observed}\nonumber\\
&\equiv& 1\;.
\label{probdefs}
\end{eqnarray}
We have simplified the last two of these distributions.  The first,
$P[{\vec h}(\lambda)]\equiv p_0(\lambda)$, follows because by
assumption the waveform depends only on $\lambda$.  This is no doubt
an oversimplification --- other parameters, such as the mass ratio of
the binary and its spins, are likely to have an important influence as
well.  To keep this analysis simple, we will focus on a single
parameter model for the merger; this is adequate for now, and
perfectly formulated for the toy model presented in Sec.\
{\ref{sec:toy}}.  The second simplification, $P({\vec g}) = 1$,
follows because ${\vec g}$ has already been measured and is therefore
determined.  Here we depart a bit from the analysis of Ref.\
{\cite{finn92}} --- Finn does {\it not} assume that this particular
datastream $\vec g$ has already been measured.  His ${\vec g}$ is {\it
undetermined}, and has a non-trivial probability distribution.

The distribution $P[{\vec g}|{\vec h}(\lambda)]$ used here is
identical to that used by Finn (modulo the extension to multiple
detectors).  It answers the question ``What is the probability that,
if the data contains the GW ${\vec h}(\lambda)$, noise is present such
that ${\vec n} + {\vec h}(\lambda)$ is consistent with the observed
data stream ${\vec g}$?''  This shows that $P[{\vec g} | {\vec
h}(\lambda)]$ is equivalent to $P[{\vec g} - {\vec h}(\lambda) | 0]$,
the probability that ${\vec g} - {\vec h}(\lambda)$ is pure noise.
Under the assumption that the noise is Gaussian, this is
{\cite{finn92}}
\begin{equation}
P[{\vec g}|{\vec h}(\lambda)] = {\cal K}
\exp\left[-{1\over2}\left({\vec g} - {\vec h}(\lambda)|
{\vec g} - {\vec h}(\lambda)\right)\right]\;,
\end{equation}
where the inner product is given by
\begin{equation}
\left({\vec a} | {\vec b}\right) \equiv 4\,{\rm Re} \int_0^\infty 
{\tilde a}_j(f)^*\left[{\bf S}_h(f)^{-1}\right]^{jk}{\tilde
b}_k(f)\,df\;,
\label{eq:innerprod}
\end{equation}
and ${\cal K}$ is a normalization constant.  Here, ${\tilde a}_j(f)$
is the Fourier transform of $a_j(t)$:
\begin{equation}
{\tilde a}_j(f) = \int_{-\infty}^{\infty} a_j(t) e^{2\pi i f t}\,dt\;,
\label{eq:fourier}
\end{equation}
and the asterisk denotes complex conjugation.  The matrix ${\bf
S}_h(f)$ is a generalization of the (one-sided) strain noise spectral
density.  Diagonal components ${\bf S}_h(f)_{jj}$ represent the usual
spectral density for detector $j$; off-diagonal elements represent
correlations between detectors $j$ and $k$.  For more details, see
Appendix A of Ref.\ {\cite{bbhII}}.  Combining these distributions,
using ${\vec g} = {\vec h}(\hat\lambda) + {\vec n}$, and defining
$\delta{\vec h}\equiv {\vec h}({\hat\lambda}) - {\vec h}(\lambda)$
yields
\begin{equation}
P[{\vec h}(\lambda) | {\vec g}] = {\cal K}p_0(\lambda)
\exp\!\left[-{1\over 2}(\delta{\vec h} | \delta{\vec h}) -
({\vec n}|\delta{\vec h}) - {1\over2}({\vec n} | {\vec n})\right].
\label{eq:probability1}
\end{equation}

Equation (\ref{eq:probability1}) depends on the specific noise
instance ${\vec n}(t)$ present in the detector network.  We circumvent
this by taking a ``frequentist'' viewpoint and ensemble averaging Eq.\
(\ref{eq:probability1}) with respect to the noise.  The resulting
distribution is our posterior distribution for $\lambda$, given data
containing $\hat\lambda$:
\begin{eqnarray}
P_{\rm post}(\lambda|\hat\lambda) &\equiv& p_0(\lambda)
{\rm E}\left\{P[{\vec h}(\lambda) | {\vec g}]\right\}_{{\vec n}}
\nonumber\\
&=& p_0(\lambda) \int P({\vec n})\,
P[{\vec h}(\lambda) | {\vec g}]\,{\cal D}{\vec n}
\nonumber\\
&=& {\cal K} p_0(\lambda)
\int{\cal D}{\vec n}\,
\exp\left[-{1\over2}\left({\vec n} | {\vec n}\right)\right]
\exp\left[-{1\over2}\left(\delta{\vec h} | \delta{\vec h}\right)
-\left({\vec n} | \delta{\vec h}\right) - {1\over2}\left({\vec n} |
{\vec n}\right)\right]\;.
\label{eq:POSTPROBINT}
\end{eqnarray}
The final integral is a functional integral, taken over all possible
network noise instances ${\vec n}$.  We evaluate it in Appendix
{\ref{app:pathintegral}}; the result is
\begin{equation}
P_{\rm post}(\lambda|\hat\lambda) = {\cal K}p_0(\lambda)
\exp\left[-{1\over4}(\delta{\vec h}|\delta{\vec h})\right].
\label{eq:postprob}
\end{equation}
This distribution is normalized so that $\int P_{\rm
post}(\lambda|\hat\lambda)d\lambda = 1$.  The normalization constant
${\cal K}$ thus depends on $\hat\lambda$ and the noise
characteristics.

As discussed in the Introduction, $P_{\rm post}({\lambda} |
{\hat\lambda})$ is not useful for designing a detector network since
it is constructed after measurement.  This is reflected by its
implicit dependence on $\hat\lambda$.  A more useful quantity is
obtained by marginalizing on ${\hat\lambda}$.  This gives a
distribution which represents the posterior distribution that we {\it
anticipate} will describe the $\lambda$ after measurement, modulo our
ignorance about $\hat\lambda$:
\begin{equation}
P_{\rm ant}(\lambda) = \int p_0(\hat\lambda)\,
P_{\rm post}(\lambda|\hat\lambda)\,d\hat\lambda.
\label{eq:antprob}
\end{equation}
The anticipated distribution answers the question ``How is $\lambda$
likely to be distributed given our ignorance about the neutron star
merger waveform?''

\subsection{Tuning the detector network}
\label{subsec:design}

We now choose the detector configuration that we anticipate will
measure $\lambda$ as accurately as possible.  Consider the mean and
variance of the distribution (\ref{eq:antprob}):
\begin{eqnarray}
\bar\lambda_{\rm ant} &=& \int \lambda\,P_{\rm ant}(\lambda)\,
d\lambda,\label{eq:mean}\\
\sigma_{\rm ant}^2 &=& \int (\lambda-\bar\lambda_{\rm ant})^2
P_{\rm ant}(\lambda)\,d\lambda,\nonumber\\
&=& \int\lambda^2P_{\rm ant}(\lambda)\,d\lambda-
\bar\lambda_{\rm ant}^2.
\label{eq:variance}
\end{eqnarray}
The mean is indicative of the typical value of $\lambda$ that a
particular observation will measure.  Likewise, the variance
(\ref{eq:variance}) is an indicator of how accurately $\lambda$ will
be measured; we anticipate that a measurement will typically find
$\lambda$ within $\sigma_{\rm ant}$ of $\bar\lambda_{\rm ant}$.

Through the inner product (\ref{eq:innerprod}), the variance
(\ref{eq:variance}) depends on the characteristics of the detector
network.  Some of these characteristics will depend on adjustable
parameters, such as the central frequencies of narrow-band detectors.
Let the $k$th such adjustable parameter be $\mu_k$.  The network which
minimizes $\sigma_{\rm ant}$ is the one which we anticipate will
measure $\lambda$ most accurately:
\begin{equation}
\min_{\mu_k} \sigma_{\rm ant}^2 \mapsto \mbox{The optimally tuned
network}\;.
\end{equation}
We call the process of finding the parameters which minimize
$\sigma_{\rm ant}^2$ ``tuning''; the network so produced is
``optimally tuned'' or simply ``optimal''.

Because of the non-Gaussian nature of the relevant distributions, a
variance such as $\sigma_{\rm ant}^2$ is a less than ideal measure of
error.  If we were truly interested in characterizing the error of a
measurement, we would instead study contours of a likelihood function
(see, e.g., Refs.\ {\cite{finn92,finnchernoff}}).  However,
$\sigma_{\rm ant}^2$ is very easy to compute, and has the correct
qualitative behavior --- it will get smaller as more rigorous measures
of error decrease.  Tuning our network in such a way that $\sigma_{\rm
ant}^2$ is minimized will drive more rigorous notions of error to be
small as well.

Since $P_{\rm post}(\lambda|\hat\lambda)$ is built from the measured
data, it depends on the measurement's signal-to-noise ratio (SNR).
Thus the anticipated probability, and the optimal network it produces,
likewise depend on SNR.  This is arguably a weakness of this
prescription for network tuning --- the network found by minimizing
$\sigma_{\rm ant}^2$ depends explicitly upon the anticipated strength
of the waves that we hope to measure.  For our present purposes, we
believe that leaving the SNR in the algorithm as an adjustable
parameter is actually desirable.  Properly eliminating the SNR
dependence would require integrating it out with a correctly
constructed probability function for its distribution.  Although there
exist estimates of the compact binary coalescence rate which we could
use to construct such a function, these estimates are currently rather
uncertain {\cite{bkb2001,kkl2002}}.  By the time that the
interferometer configurations we consider here can be installed,
LIGO-type detectors could be detecting many events per year, or just a
few.  If many are detected, then observers may choose to focus their
merger analyses on the strongest events in this large set.  On the
other hand, if the rate turns out to be low, observers will be forced
to work with relatively weak signals.  It is therefore useful to see
how SNR choice affects network design, exploring how tuning will work
as a function of the SNR levels that turn out to be appropriate.
Accordingly, we choose to leave in the SNR as a parameter for now.

Finally, note that the prior probability $p_0(\lambda)$ plays a
crucial role, through Eqs.\ (\ref{eq:postprob}) and
(\ref{eq:antprob}), in determining the optimal detector network.  This
gives a natural way to update the network as our knowledge of merger
waves improves: we continually update the priors as we learn more
about $\lambda$, and concomitantly update the network to zoom-in on
the true value of $\lambda$.  This updating is the key element of the
network tuning algorithm.

Explicitly, this algorithm works as follows:

\begin{enumerate}

\item Before the first merger is measured, our understanding of
$\lambda$ is likely to be crude.  Take the prior distribution to be
uniform between some upper and lower limits: $p_0(\lambda)
=\mbox{const}$, $\lambda_{\rm MIN}\le\lambda\le\lambda_{\rm MAX}$.

\item Configure the detector network by minimizing Eq.\
(\ref{eq:variance}) with this uniform prior.

\item Measure a merger.  Build the posterior distribution
$P_{\rm post}(\lambda | {\hat\lambda})$ [Eq.\ (\ref{eq:postprob})]
from the measured data.

\item Use this posterior distribution as the prior for the next
measurement: set $p_0^{\rm new}(\lambda)= P_{\rm
post}(\lambda|\hat\lambda)$.

\item Reconfigure the network by minimizing (\ref{eq:variance}) with
this prior.

\item Go to 3.

\end{enumerate}

As we will show in Sec.\ {\ref{sec:results}}, this algorithm
effectively zooms in on the true parameterization $\hat\lambda$ of the
waves in the datastream, at least for the toy model we present next.

\section{Toy model}
\label{sec:toy}

Here we describe the simplified model we use for the waveform (Sec.\
{\ref{subsec:wave}}) and for the detector noise (Sec.\
{\ref{subsec:detectornoise}}).  These models make it possible to
evaluate the posterior distribution analytically (Sec.\
{\ref{subsec:postprobevaluate}}).  We discuss the shortcomings of this
toy description in Sec.\ {\ref{subsec:shortcomings}}.  We plan to
address these shortcomings in future work.

\subsection{Waveform}
\label{subsec:wave}

The gravitational waveform has two polarizations, $h_+$ and $h_\times$
(named for the axes of force lines associated with them).  Each
detector in the network measures some linear combination of these
polarizations:
\begin{equation}
h_j = F_{j,+} h_+ + F_{j,\times} h_\times.
\end{equation}
The functions $F_{j,+,\times}$ depend on the type of GW detector used
for the measurement.  For interferometers, standard formulae for
$F_{+,\times}$ are given in, e.g., Ref.\ {\cite{300yrs}} [Eqs.\
(104a,b) and preceding text].  For certain spherical acoustic
detectors (cf.\ Ref.\ {\cite{blair}} and references therein),
$F_+\simeq F_\times\simeq {\rm const}$ --- spherical antennae have
nearly equal sensitivity to both polarizations, independent of the
source's sky position.

The polarizations of the leading quadrupole harmonic of the inspiral
waveform are, in the frequency domain,
\begin{eqnarray}
{\tilde h}_+(f) &=& 2(1 + \cos^2\iota) {\tilde h}_Q(f),
\nonumber\\
{\tilde h}_\times(f) &=& 4\cos\iota\,e^{i\pi/2}\,{\tilde
h}_Q(f);
\nonumber\\
{\tilde h}_Q(f) &\equiv& \sqrt{5\over96} {\pi^{-2/3}{\cal M}^{5/6}
\over r}f^{-7/6} e^{i\Phi(f)}.
\label{eq:quadrupole}
\end{eqnarray}
The angle $\iota$ is between the line of sight to the binary and its
orbital angular momentum, $r$ is the distance to the source, ${\cal
M}\equiv (m_1 m_2)^{3/5}/(m_1+m_2)^{1/5}$ is the binary's ``chirp
mass'', and $\Phi(f)$ is a phase function, described further below.

To make our toy model for the waveform, we first simplify this
inspiral in two ways.  First, we restrict ourselves to equal mass
binaries: we set $m_1 = m_2 = 1.4\,M_\odot$.  Since we are focusing on
neutron star binaries, this is not a drastic simplification.  Second,
we use only the Newtonian, quadrupole contribution to the phase,
ignoring higher-order post-Newtonian contributions:
\begin{equation}
\Phi(f) = 2\pi f t_c - \phi_c - {\pi\over4} + {3\over4}
\left(8\pi{\cal M}f\right)^{-5/3}\;,
\label{eq:NQphase}
\end{equation}
where $t_c$ is the time at which the frequency (formally) diverges to
infinity, and $\phi_c$ is the phase at that time.

To represent the transition from inspiral to merger, we multiply
$h_{+,\times}$ by $\Theta(\lambda f_{\rm kHz} - f)$, where $\Theta$ is
the step function and $f_{\rm kHz} = 1000\,\mbox{Hz}$.  We assume that
$\lambda$ lies between $0.5$ and $1.5$; this is consistent with the
results seen in Ref.\ {\cite{fgrt}}.  This extremely simple waveform
captures the essence of coalescence models in which the GWs shut off
very rapidly as the neutron stars come into contact with one another.
Because we neglect higher order contributions to the phasing, and
because our model shuts off the wave emission far more rapidly than is
seen in any actual calculation, our results should be considered
illustrative rather than definitive.

The frequency domain GW signal measured in detector $j$ of the network
is therefore
\begin{equation}
{\tilde h}_j(f) = \left[2(1+\cos\iota^2)F_{j,+} + 4\cos\iota
\,e^{i\pi/2}\,F_{j,\times}\right]
{\tilde h}_Q(f)\Theta(\lambda f_{\rm kHz} - f)\;.
\label{eq:toywaveform}
\end{equation}
A very useful quantity which we can calculate from this waveform is
the SNR $\rho_j$:
\begin{eqnarray}
\rho^2_j &=& 4\int_0^\infty {|{\tilde h}_j(f)|^2\over S_h(f)_j}df
\label{eq:snr_def}\\
&=& {5|\Psi|_j^2\over24} {\pi^{-4/3}{\cal M}^{5/3}\over r^2}
\int_0^{\lambda f_{\rm kHz}} {f^{-7/3}\over S_h(f)_j}df\;.
\label{eq:snr_toy}
\end{eqnarray}
The first line is a standard formula for the SNR of a GW measurement
(cf.\ Refs.\ {\cite{finn92,finnchernoff,cf}}).  The quantity
$S_h(f)_j$ is the spectral density of noise in the $j$-th detector; we
discuss our toy description of this quantity in the following section.
In the second line, we have specialized to the toy waveform.  We have
introduced here the angular function
\begin{equation}
|\Psi|_j^2 = 4\left[(1 + \cos^2\iota)^2 F_{j,+}^2 + 4\cos^2\iota
 F_{j,\times}^2\right]\;.
\label{eq:Psidef}
\end{equation}
Note that $|\Psi|_j^2$ is identical to the $\Theta^2$ defined in Ref.\
{\cite{finnchernoff}}; we use a slightly different notation to avoid
confusion with the step function.  This function lies in the range $0
\le |\Psi|_j^2 \le 16$; its value averaged over all sky positions and
source orientation angles is $\overline{|\Psi|_j^2} = 2.56$.  The
distribution of $|\Psi|_j^2$ is quite asymmetric, with values smaller
than $\overline{|\Psi|_j^2}$ more likely than larger ones.  See Ref.\
{\cite{finnchernoff}} for further discussion.

We plot a representation of this waveform in Fig.\ {\ref{fig:noise}},
comparing it to our model of the detector noise (discussed in detail
in the following section).  The waveform plotted here is somewhat
massaged --- we actually plot
\begin{equation}
h_{\rm plot}(f) = \sqrt{f |{\tilde h}(f)|^2}\;.
\end{equation}
This quantity has the same units as the noise spectrum, and has the
nice feature that the integral of $h_{\rm plot}(f)^2$ over the noise
is the squared signal-to-noise ratio $\rho^2$: using Eq.\
(\ref{eq:snr_def}), we see that
\begin{equation}
\rho^2_j = 4\int d\ln f\,{h_{\rm plot}(f)_j^2\over S_h(f)_j}\;.
\end{equation}
The waveform we show corresponds to a pair of $1.4\,M_\odot$ neutron
stars with optimal orientation ($|\Psi|^2 = 16$) at 500 Mpc; or, with
``average orientation'' ($|\Psi|^2 = 2.56$) at 200 Mpc.  We have
chosen $\lambda = 0.8$, so that the waveform's merger cutoff is at 800
Hz.

Inserting the toy waveform (\ref{eq:toywaveform}) into Eq.\
(\ref{eq:postprob}) yields
\begin{eqnarray}
P_{\rm post}(\lambda|\hat\lambda) &=& p_0(\lambda)
\exp\left[-{\rm Re} \int_0^\infty
[2(1+\cos\iota^2)F_{j,+} + 4\cos\iota\,e^{i\pi/2}\,
F_{j,\times}][2(1+\cos\iota^2)F_{k,+}\right.\nonumber\\
&+&\left. 4\cos\iota\,e^{i\pi/2}\,F_{k,\times}]
\left[\Theta(\hat\lambda f_{\rm kHz} - f)
-\Theta(\lambda f_{\rm kHz} - f)\right]^2 |{\tilde h}_Q(f)|^2
\left[{\bf S}_h(f)^{-1}\right]^{jk} df \right]\;.
\label{eq:probability4}
\end{eqnarray}
Define $\lambda_{\rm LO}\equiv\min(\hat\lambda,\lambda)$,
$\lambda_{\rm HI}\equiv\max(\hat\lambda,\lambda)$.  Because of the
$\Theta$ functions in Eq.\ (\ref{eq:probability4}), the domain of
integration is restricted to the range $f_{\rm LO}\le f\le f_{\rm HI}$
(where $f_{\rm LO} = \lambda_{\rm LO}f_{\rm kHz}$, $f_{\rm HI}
=\lambda_{\rm HI}f_{\rm kHz}$).  For the range of $\lambda$ that we
have selected, the detector noise in the merger band is dominated by
laser shot noise.  We will use this to simplify Eq.\
(\ref{eq:probability4}) further in what follows.

\subsection{Detector noise}
\label{subsec:detectornoise}

Very detailed parameterized models for the LIGO noise budget exist and
can be manipulated using codes such as {\sc bench} {\cite{bench}}
(which was used for the analysis of {\cite{hhs}}).  For our purposes,
much simpler descriptions that are simple to manipulate analytically
will be mostly adequate.

We will take our broad-band detectors to have the noise curve
anticipated for the second-generation LIGO detectors (``LIGO-II
noise''), as described in Refs.\ {\cite{whitepaper,hhs}}.  In
particular, since we focus on measurements of binary neutron star
inspiral, we will assume the broad-band measurements are made using
the interferometer configuration optimized for that source (the curve
labeled ``NS-NS'' in Figure 1 of Ref.\ {\cite{whitepaper}}); we show
it in Fig.\ {\ref{fig:noise}}.

The NS-NS LIGO-II noise profile we use is a broad-band signal-recycled
configuration.  In the shot noise dominated band, its noise spectrum
grows with an $f^2$ law:
\begin{eqnarray}
S^{\rm BB}_h(f) &=& \gamma_{\rm NS-NS}f^2
\nonumber\\
&=& \left(1.5\times10^{-52}\,\mbox{Hz}^{-3}\right)f^2\;.
\label{eq:bbnoise}
\end{eqnarray}
The numerical value of the parameter $\gamma_{\rm NS-NS}$ given here
was found by fitting to the high-frequency end of the NS-NS LIGO-II
noise curve given in {\cite{whitepaper}}.

We will take our narrow-band detectors to be LIGO-II detectors
operated in a narrow-band configuration; cf.\ discussion in Refs.\
{\cite{whitepaper,hhs}}.  A detailed description of the noise in this
configuration can be found using {\sc bench}; near the most sensitive
frequency of such a configuration, it is well described by the
approximate formula {\cite{300yrs,klm,krolak}}
\begin{equation}
S^{\rm NB}_h(f) \simeq 2S_0\left({\Delta f\over f_0}\right)\left[1 + 
4\left(f - f_R\over\Delta f\right)^2\right]\;.
\label{eq:nbnoise1}
\end{equation}
The parameters $S_0$ and $f_0$ appearing here are
\begin{eqnarray}
S_0 &=& {\hbar \lambda_L c\over4\pi \eta I_0}\left(\beta\over L\right)^2
\simeq2.1\times 10^{-51}\,\mbox{Hz}^{-1}\;,
\label{eq:S0num}\\
f_0 &=& {\beta c\over4\pi L}\simeq 0.2\,\mbox{Hz}\;.
\label{eq:f0num}
\end{eqnarray}
The numerical values of the laser wavelength $\lambda_L$, laser power
$I_0$, photodiode efficiency $\eta$, mirror power loss $\beta$, and
armlength $L$ are given in Table II of Ref.\ {\cite{hhs}}.  The
frequency $f_R$ is controlled by the position of one of the mirrors in
the narrow-banded interferometer, and as such is not too difficult to
adjust; it might even be possible to {\it dynamically} adjust this
position, in order to track the evolution of a particularly
interesting feature.

The bandwidth $\Delta f$ is rather more complicated.  It is largely
set by losses in the mirrors, but also depends in an important way on
the position of the mirror that controls $f_R$.  Hence, as we tune
$f_R$, the bandwidth changes as well.  For the purpose of our toy
model, we will ignore this effect; examples of the toy narrow-band
noise spectrum (with fixed bandwidth) are shown in Fig.\
{\ref{fig:noise}}.  We emphasize that the bandwidth {\it cannot} be
treated as constant in a realistic analysis.  Indeed, there are many
coupled effects that appear when a realistical description of a
narrow-band interferometer is used.  For example, configurations that
improve detector response in a narrow-band at high frequency can
actually improve response at low frequency as well
{\cite{ds_private,bc2001}}.

For the toy model, we will approximate the narrow-band detectors as
having infinitesimal bandwidth:
\begin{equation}
{1\over S^{\rm NB}_h(f)}\simeq {\pi f_0\over4 S_0}\delta(f - f_R)\;.
\label{eq:nbnoise}
\end{equation}
This form is particularly easy to use, and is not a bad description of
the interferometer response around the center frequency when the
bandwidth is very narrow.  The prefactor is chosen so that the
integral of $1/S^{\rm NB}_h(f)$ is the same for Eqs.\
(\ref{eq:nbnoise1}) and (\ref{eq:nbnoise}).  We will treat the
frequency $f_R$ of each narrow-band detector in our network as an
adjustable parameter, and tune our network by varying those
frequencies to find the optimal network for measuring merger waves.
Note that by using such a narrow detector response, we throw away
potentially useful information.  This is an obvious point of
improvement for an analysis that goes beyond this toy model.

\subsection{Evaluation of the posterior probability}
\label{subsec:postprobevaluate}

We now use the toy description of the noise to evaluate analytically
the posterior probability distribution (\ref{eq:probability4}).
First, we assume that the various individual detectors on the network
have uncorrelated noises: we set the off-diagonal terms in the matrix
${\bf S}_h(f)^{-1}$ to zero.  This should be a reasonable assumption
in the high frequency, shot-noise-dominated regime appropriate to this
analysis --- each interferometer will have its own laser and optics
system, and should therefore be reasonably isolated from all other
detectors (though events affecting interferometers at a common site
could violate this assumption).  Equation (\ref{eq:probability4}) then
simplifies to
\begin{equation}
P_{\rm post}(\lambda|\hat\lambda) = p_0(\lambda)
\exp\left[-\sum_{j = 1}^N |\Psi|_j^2
\int_{f_{\rm LO}}^{f_{\rm HI}}
{|{\tilde h}_Q(f)|^2\over S_h(f)_j} df \right]\;,
\label{eq:probability5}
\end{equation}
where $N$ is the total number of detectors, broad-band plus
narrow-band.

To simplify this analysis further, we will assume that all detectors
are housed within a single facility, so that the functions
$|\Psi|_j^2$ take a single value $|\Psi|^2$ for all $j$.  Further, we
will assume there is only one broad-band detector in this facility,
and $N_N$ narrow-band detectors.  The sum in the exponent of Eq.\
(\ref{eq:probability5}) can now be simplified:
\begin{equation}
\sum_{j = 1}^N |\Psi|_j^2
\int_{f_{\rm LO}}^{f_{\rm HI}}
{|{\tilde h}_Q(f)|^2\over S_h(f)_j} df = |\Psi|^2
\int_{f_{\rm LO}}^{f_{\rm HI}}
{|{\tilde h}_Q(f)|^2\over S^{\rm BB}_h(f)} df
+|\Psi|^2\sum_{j = 1}^{N_N}\int_{f_{\rm LO}}^{f_{\rm HI}}
{|{\tilde h}_Q(f)|^2\over S^{\rm NB}_h(f)_j} df\;.
\label{eq:exponentsum}
\end{equation}
Using Eqs.\ (\ref{eq:bbnoise}) and (\ref{eq:nbnoise}), these integrals
are simple to evalute.  The result is
\begin{eqnarray}
P_{\rm post}(\lambda|\hat\lambda) &=& p_0(\lambda)
\exp\left[-{3\over20}{\rho_{\rm insp}^2\over\sigma_7}
{1\over\gamma_{\rm NS-NS}} f_{\rm kHz}^{-10/3}
\left|{1\over\hat\lambda^{10/3}} - {1\over\lambda^{10/3}}\right|
\right.\nonumber\\
&-&\left.{\pi\over16}{\rho_{\rm insp}^2\over\sigma_7}{f_0\over S_0}
f_{\rm kHz}^{-7/3}
\sum_{j=1}^{N_N}\left({f_{\rm kHz}\over f_{R,j}}\right)^{7/3}
\left|\Theta(\hat\lambda f_{\rm kHz} - f_{R,j}) -
\Theta(\lambda f_{\rm kHz} - f_{R,j})\right|\right]\;.
\label{eq:toypostprob}
\end{eqnarray}
The frequency $f_{R,j}$ is the adjustable frequency of peak
sensitivity for the $j$-th narrow-band detector.  The step functions
enforce the fact that the integral over the narrow-band noise is
non-zero only if $f_{\rm LO} < f_{R,j} < f_{\rm HI}$.  We have
simplified this expression considerably by normalizing the amplitude
of the GWs to the inspiral SNR $\rho_{\rm insp}$ measured in the
broad-band detector.  Using Eq.\ (\ref{eq:snr_toy}),
\begin{eqnarray}
\rho^2_{\rm insp} &\simeq& {5|\Psi|^2\over24}{\pi^{-4/3}{\cal
M}^{5/3}\over r^2} \int_0^\infty {f^{-7/3}\over S^{\rm BB}_h(f)}df
\nonumber\\
&=& {5|\Psi|^2\over24}{\pi^{-4/3}{\cal M}^{5/3}\over r^2}\sigma_7\;.
\label{eq:BBSNR}
\end{eqnarray}
Taking the upper limit of the integral to infinity (rather than
$\lambda\times 1000\,{\rm Hz}$) is fine for the inspiral SNR since
most of the signal accumulates near $f \sim 100$ Hz --- the upper
frequency has little effect.  With the broad-band LIGO-II noisecurve
used here, $\sigma_7 = 8.3\times 10^{44}\,\mbox{Hz}^{-1/3}$.  For a
binary consisting of a pair of $1.4\,M_\odot$ neutron stars, the SNR
becomes
\begin{equation}
\rho_{\rm insp} = 11.7\left({|\Psi|^2\over2.56}\right)^{1/2}
\left({200\,{\rm Mpc}\over r}\right)\;.
\label{eq:BBSNR2}
\end{equation}
With this parameterization, we see how the tuning algorithm will
depend upon SNR, as described in Sec.\ {\ref{subsec:design}}.  Note
that, in this toy model, the inspiral SNR and the frequency $f_{R,j}$
completely determine the SNR measured in the $j$-th narrow-band
detector:
\begin{eqnarray}
\rho^2_{{\rm NB},j} &=& 4\int_0^\infty {|{\tilde h}_Q(f)|^2\over
S^{\rm NB}_h(f)_j} df
\nonumber\\
&=& {\pi\over16}{\rho^2_{\rm insp}\over\sigma_7}{f_0\over S_0}
f_{R,j}^{-7/3}\;.
\label{eq:NBSNR}
\end{eqnarray}
Plugging in the values of $f_0$, $S_0$, and $\sigma_7$, we find
\begin{equation}
{\rho_{\rm NB}\over\rho_{\rm insp}} = 0.05\left({1000\over
f_{R,j}}\right)^{7/6}\;.
\label{eq:SNRratio}
\end{equation}
The narrow-band SNR drops off very quickly with frequency.  This
follows directly from our toy model for the waveform --- the amplitude
decays with frequency as $f^{-7/6}$ up to the step function break, so
the SNR decays likewise.

The anticipated probability distribution is found by plugging
(\ref{eq:toypostprob}) into (\ref{eq:antprob}).  It is best evaluated
numerically.

\subsection{Shortcomings of this toy description}
\label{subsec:shortcomings}

Although useful for demonstrating how network tuning works, we cannot
pretend that the toy model is anything more than illustrative.  First,
the waveform we use [Eq.\ (\ref{eq:quadrupole})] is drastically
oversimplified.  Although it produces an energy spectrum qualitatively
similar to that seen in recent computations (e.g., Ref.\
{\cite{fgrt}}), it shuts off the inspiral waves far more quickly than
those calculations predict.  In the spectra developed in Ref.\
{\cite{fgrt}}, the GW energy emission drops from the (relatively
large) level of the quadrupole inspiral by about an order of magnitude
over a band of a few hundred Hz.  By not rolling off more gradually,
the toy waves are likely too large as the merger frequency $\lambda
f_{\rm kHz}$ is approached.  Second, the infinitesimal bandwidth
(\ref{eq:nbnoise}) we use for the narrow-band detectors does not
describe the configurations that will be used in realistic second
generation interferometers --- the actual detectors will have some
finite bandwidth of tens to hundreds of Hz.

The toy model captures enough of the ``feel'' of the real problem that
we are confident this analysis should be a useful demonstration of how
network tuning will work with realistic waves and detectors.
Nonetheless, because of the details that have been thrown away, we
emphasize that this analysis must be followed up by calculations that
are not oversimplified.  The interplay between these neglected
features is likely to be rather important --- for instance, the finite
bandwidth of the real narrow-band detectors may help to detect the
relatively slow shut off that may describe merger waves in nature.
This would also be an opportunity to test out ideas such as dynamical
tuning of narrow-band detectors.

\section{Results}
\label{sec:results}

We now test the network tuning algorithm on the toy model.  We
implement tuning by numerically finding the minima of the anticipated
variance $\sigma_{\rm ant}^2$ [which is itself built from a numerical
construction of the antipicated distribution, using Eq.\
(\ref{eq:toypostprob}) in Eq.\ (\ref{eq:antprob})].  This was done
with Powell's multidimensional line minimization technique, using
Brent's method for the successive line minimizations (cf.\ Ref.\
{\cite{recipes}}).  This approach works very well at finding local
minima.  If the number of minima is not too large, one can simply
compute the set of all local minima, then find the global minimum of
$\sigma_{\rm ant}^2$ in that set.  This is adequate for the toy model
provided that the number of narrow-band detectors is not too large.
It may not work nearly so well when more realistic waveforms and noise
models are used; robust global minimization methods (e.g., simulated
annealing) might work better in this case.

Before examining merger measurements in detail, we discuss at some
length the detector networks that tuning produces
(Sec. {\ref{subsec:networks}}).  The toy model is simple enough that
we can easily understand why this algorithm tunes the network to the
parameters that it chooses; we expect that the intuition developed
here should apply more generally.  We then look at merger measurement
in Sec.\ {\ref{subsec:measure}}.  We begin with a uniform prior
distribution for $\lambda$ (imagining that, aside from an upper and a
lower bound, we are completely ignorant of $\lambda$'s value) and show
that repeated measurements of NS-NS merger drive the distribution to
gradually peak on the true value $\hat\lambda$.  We look at this
process at two SNR levels ($\rho_{\rm insp} = 10$ and $30$), and
consider measurement using 1 and 2 narrow-band detectors.

\subsection{Optimal detector networks}
\label{subsec:networks}

Since our network tuning procedure works by minimizing $\sigma_{\rm
ant}^2$, it is useful to understand this quantity's properties.
Figure {\ref{fig:var}} shows the anticipated variance when
measurements are made with a single narrow-band detector.  We assume a
uniform prior distribution for $\lambda$ between $\lambda = 0.5$ and
$\lambda = 1.5$, and show results for inspiral SNR $\rho_{\rm insp} =
10$ (upper panel) and $90$ (lower panel).

In both cases, we see two minima: one is in the vicinity of $600 -
800$ Hz, the other is near $1200 - 1400$ Hz.  For $\rho_{\rm insp} =
10$, the lower frequency minimum is the global minimum; for $\rho_{\rm
insp} = 90$, the reverse holds.  This interesting result has a simple
explanation.  As we move to higher frequencies, the broad-band
detector's shot noise grows increasingly more important, with a
spectrum proportional to $f^2$.  This makes it advantageous to place a
narrow-band detector at high frequency, where it can compensate for
the broad-band detector's degrading performance.  However, our assumed
waveform (\ref{eq:toywaveform}) gets very weak at high frequencies:
the amplitude (prior to the cutoff enforced by the step function)
decays with frequency as $f^{-7/6}$.  The SNR measured by a
high-frequency narrow-band detector is potentially very small [cf.\
Eq.\ (\ref{eq:SNRratio})], and we are more likely to get a ``false
positive'' --- an incorrect signal measurement due to noise at high
frequencies.  These two phenomena compete.  Which of these two
dominates in the end depends upon the overall signal amplitude (here
set by the inspiral SNR).

At low SNR, the need to avoid false positives tends to win out --- the
low frequency minimum at $f\sim 630$ Hz is the global minimum of
$\sigma_{\rm ant}^2$.  As the signal strength is increased, the danger
of false positives decreases.  At $\rho_{\rm insp} = 90$, the tuning
algorithm places the narrow-band detector at $f\sim 1380$ Hz --- the
signal is strong enough that the tuning algorithm chooses to
compensate for the degraded sensitivity of the broad-band detector.

This pattern of minima location for uniform prior distribution --- a
high frequency minimum near $f\sim 1200 - 1400$ Hz, a low frequency
minimum near $f\sim 600 - 800$ Hz --- holds as we increase the number
of narrow-band detectors.  Table {\ref{tab:network}} shows the
narrow-band detector frequencies that minimize $\sigma_{\rm ant}^2$
for inspiral SNR 10, 30, and 90.  Note that for $\rho_{\rm insp} =
10$, the tuning algorithm {\it always} places the narrow-band
detectors at low frequency, even for $N_N = 5$.  At higher SNR, it
tends to distribute the detectors more evenly.

It's worth noting at this point that we assume the noise statistics of
all our detectors is Gaussian, and hence our discussion of the
likelihood of ``false positives'' is based on what may be a somewhat
optimistic assumption of how the detectors will behave.  (Since we
assume that an inspiral signal has been detected and that the merger
waves rapidly follow, Gaussianity may not be terribly optimistic ---
non-Gaussian events which violate this assumption will have to occur
in a very narrow, specific time window.)  Using multiple detectors in
a similar band has the additional benefit of serving as checks on one
another; it is difficult to know how one's instruments will behave
until they have been studied and their statistical behavior is known.

\subsection{Measurement sequences}
\label{subsec:measure}

Next we examine measurement sequences, following the algorithm
outlined at the end of Sec.\ {\ref{sec:formal}}.  As already
mentioned, the key element of this algorithm is updating the priors
after each measurement: we replace the prior distribution
$p_0(\lambda)$ with the posterior distribution built from the measured
data, $P(\lambda|\hat\lambda)$.  In this simulated sequence of
measurements, we assume for simplicity that each measurement in the
sequence has identical SNR; we also (as discussed in Sec.\
{\ref{sec:formal}}) use a posterior distribution that has been
ensemble averaged with respect to the noise.  A more realistic
calculation would randomly pick a noise realization from the
distribution of possible noise functions, and would allow the SNR to
vary from measurement to measurement in accordance with estimates of
the distribution of binary inspiral events.

We first look at how well detector networks can measure $\lambda$ for
moderate inspiral SNR ($\rho_{\rm insp} = 10$) and with a single
narrow-band detector.  We show a measurement sequence for $\hat\lambda
= 0.8$ in Fig.\ {\ref{fig:snr10_nb1_lh0.8}}.  The upper left hand
panel shows the posterior probability that would be obtained if no
narrow-band detectors were used for the measurement.  It is
practically flat --- there is essentially no information about
$\lambda$ in this distribution.  The other panels show how the
distribution evolves as multiple measurements are made using a
narrow-band detector.  The distribution gradually becomes peaked as
more and more measurements are made.  Network tuning ``zooms-in'' on
the true merger parameterization $\hat\lambda$, placing the
narrow-banded detector where it most effectively measures the merger
waves.  After 9 measurements, the distribution is clearly peaked
around $\hat\lambda$; it is markedly peaked after 15 measurements.
Such a strongly peaked distribution indicates that $\lambda$ has been
measured very accurately --- the narrowness of the distribution means
that the error in $\lambda$'s determination is very small.

When $\hat\lambda = 1.2$, it takes many more measurements before the
distributions become peaked; this is shown in Fig.\
{\ref{fig:snr10_nb1_lh1.2}}.  This is not surprising --- the merger
features are at $1200$ Hz in this case, not $800$ Hz, and so the
expected narrow-band SNR is smaller by a factor of $(800/1200)^{7/6} =
0.6$.  Because of the low SNR, the tuning algorithm tends to avoid
placing the detector at high frequencies.  The algorithm walks the
narrow-band detector up to high frequency fairly slowly, doing so at a
rate that ensures we have ruled out a low value of $\hat\lambda$
before ``testing the waters'' at high values.  After 25 measurements,
we have narrowed down to a relatively broad distribution centered at
$\lambda\sim 1.1$; after a total of 50 measurements, the distribution
is strongly peaked near $\hat\lambda$.

Figure {\ref{fig:snr10_nb2_lh1.2}} shows a sequence equivalent to that
used for Fig.\ {\ref{fig:snr10_nb1_lh1.2}}, but now using two
narrow-band detectors.  The convergence onto the true parameterization
is quite a bit quicker in this case.  With two narrow-band detectors,
the tuning algorithm has much more freedom to choose how it zooms in
on the merger waves than when only one is used --- the detectors can
be placed at both ends of the merger frequency space and slide inward,
or both can be placed at one end.  In the case shown, the distribution
found after 28 measurements with two narrow-band detectors is nearly
the same as it was after 50 measurements with one narrow-band
detector.

The posterior distribution becomes peaked near $\hat\lambda$ much more
quickly if the measurements include several events at higher SNR.
Figure {\ref{fig:snr30_nb1_lh1.2}} shows the convergence of the
distribution for $\hat\lambda$ when each measurement is at $\rho_{\rm
insp} = 30$ rather than 10.  There is much less danger of a false
positive at this SNR level, so the algorithm is able to zoom in on the
true parameterization much more aggressively.  As a consequence, the
distribution is sharply peaked after only nine measurements ---
$\lambda$ is measured quite accurately relatively quickly at this SNR.
The convergence is even quicker when $\hat\lambda = 0.8$: although we
do not show the corresponding plots, we find that the posterior
distributions become quite peaked about $\hat\lambda$ after only 4
measurements with $\rho_{\rm insp} = 30$.

\section{Conclusions and directions for further work}
\label{sec:future}

The results of the previous section show that network tuning works
effectively, at least within the context of the toy model: in all
cases, we find that after some number of measurements the probability
distribution describing what is known about $\lambda$ will become
sharply peaked about $\hat\lambda$.  The number of measurements that
are needed depends quite a bit on the particulars of the waveform and
of the detector network used.  In particular, we can draw two
conclusions that are not at all surprising:

\begin{itemize}

\item Multiple narrow-band detectors helps the measurement process
greatly.  In particular, the improvement in going from one narrow-band
detector to two detectors can be significant.

\item The rate at which measurements converge onto an accurate
description of the merger waves depends strongly on those waves'
measured SNR.  These can be seen in the drastic difference in the
convergence of measurement sequences for $\hat\lambda = 0.8$ and
$\hat\lambda = 1.2$ at fixed inspiral SNR (Fig.\
{\ref{fig:snr10_nb1_lh0.8}} versus Fig.\ {\ref{fig:snr10_nb1_lh1.2}})
and in the drastic difference seen when the wave amplitude is
increased by a factor of 3 (Fig.\ {\ref{fig:snr10_nb1_lh1.2}} versus
Fig.\ {\ref{fig:snr30_nb1_lh1.2}}).

\end{itemize}

From the first of these conclusions, we advocate investigating the
possibility of running at least two detectors in the world-wide GW
detector network in a narrow-band configuration.  For example, LIGO
already has 3 running interferometers (a 4 km broad-band detector and
a 2 km broad-band detector at Hanford, Washington, plus a 4 km
broad-band detector at Livingston, Louisiana).  Room has been made in
the facilities for additional interferometers, though cost is likely
to limit the number that can actually be installed.  If a total of 4
interferometers can be used in the LIGO facilities, it may be
worthwhile to reconfigure one of the broad-band detectors as a
narrow-band instrument in order to search for merger waves, assuming
the loss of a broad-band detector would not seriously impact other
science goals (e.g., if other detectors worldwide are able to fill the
gap).

Second, we strongly advocate continued theoretical efforts to model
and understand the properties of merger waves.  Of particular interest
are robust characteristics such as spectral breaks and features that
should be measureable without needing detailed models of the waves'
phasing.  This kind of understanding will make it possible to choose a
parameterization of the waves, similar to our parameter $\lambda$,
that leaves a strong mark on the waveform and can be measured
reasonably well.  A robust theoretical foundation for the merger waves
will make it possible to choose our priors and configure our network
usefully so that measurements will teach us about the merger waves
relatively quickly.

As we have repeatedly emphasized, this analysis should be considered a
first, proof-of-concept presentation of how network tuning can work.
We believe it is very important that, having presented the principles
and a simple example of how they work, this effort be followed by a
detailed followup analysis that uses a description of the waves and
the detectors that is more sophisticated than our toy model.  In wave
modeling, we must investigate merger waveforms that are not as
trivially simple as those used in the toy model.  In principle, this
is not too difficult even now --- some groups have already produced
examples of merger waves that show the influence of the dense matter
EOS {\cite{rs,fgrt}}.  Those waveforms could be built into this
analysis without too much difficulty.

In the detector description, it is very important that the true
bandwidth behavior of the narrow-band interferometers be included ---
our infinitesimal bandwidth description is clearly inadequate.  Real
detectors will be far more complicated than the toy description given
here, and many of these complications are actually coupled (for
example, changing the center frequency of a narrow-band detector
impacts its bandwidth).  A realistic assessment of how well network
tuning can work and how it should be implemented must take into
account these various coupled complications.  It would also be useful
to include acoustic narrow-band detectors in this analysis
{\cite{hhs}}, in order to assess what role they could play in concert
with the broad-band interferometric detectors.

It would not be too surprising if including a non-zero bandwidth
detector in this description improved the performance of network
tuning --- finite bandwidth detectors will sample a moderate span of
frequency, and can thus look for merger power more broadly than do the
toy detectors considered here.  This comes at a bit of cost: the
broader the bandwidth of the detector, the less amplitude sensitivity
it has at its center frequency.  A combination of very narrow-band and
moderately narrow-band detectors may turn out to provide the best of
both worlds.  It would also be interesting to explore how well
dynamical tuning might work: if it is possible to detect an inspiral
in real time (plausible for strong sources measured in advanced
detectors), we may be able to adjust the center frequency of the
narrow-band interferometer to follow the gravitational waves as they
evolve through the late inspiral and merger.  This could significantly
speed up convergence of the parameter distributions.

Other simplifications that we have used here should also be removed in
a follow-up analysis.  In particular, in simulating a sequence of
measurements we have assumed that all measurements are at some fixed
SNR.  This is obviously incorrect.  In setting the measured SNR, it
would be more appropriate to assume a uniform distribution in volume
out to some distance (say 500 Mpc), and to set the distance to each
source according to that distribution.  Then, each measured event
should have its angular function $\Psi$ [defined in the text following
Eq.\ (\ref{eq:probability5})] taken from the appropriate distribution
(cf.\ Ref.\ {\cite{finnchernoff}}).  Also, when we construct the
posterior distribution following a simulated measurement, it would be
much more appropriate to construct a noise instance ${\vec n}$ and
hence simulate the data stream ${\vec g}$, rather than ensemble
averaging.

We advocate undertaking this kind of follow-up analysis soon, so that
these algorithms are well-understood when it become possible to
actually construct the relevant detector networks, and so that what is
learned from them can impact the design of future detectors.

\begin{acknowledgments}

I thank Kip Thorne, who played an instrumental role in helping to
develop the major ideas that lie at the core of this analysis, and
also offered helpful comments and criticism.  I am also very grateful
to Krishna Rajagopal for useful comments on a draft of this paper, and
to Gregg Harry and David Shoemaker for adding much-needed experimental
realism to an ignorant theorist's utopian ideas.  An early version of
these ideas was first presented in Chapter 4 of the author's Ph.\ D.\
thesis (Ref.\ {\cite{phd}}), which was supported by NSF Grant
PHY-9424337 at Caltech.  This work was supported at the KITP by NSF
Grant PHY-9907949.

\end{acknowledgments}

\appendix

\onecolumngrid

\section{Evaluation of the integral (\ref{eq:POSTPROBINT})}
\label{app:pathintegral}

In this appendix, we evaluate the integral (\ref{eq:POSTPROBINT}), and
thereby calculate the posterior probability distribution $P_{\rm
post}(\lambda|\hat\lambda)$.  First, consider a single detector: the
posterior probability in this case is
\begin{equation}
P_{\rm post}(\lambda | \hat\lambda) =
{\cal K} p_0(\lambda)
\int\exp\left[-{1\over2}\left(n | n\right)\right]
\exp\left[-{1\over2}\left(\delta h|\delta h\right) -
\left(n | \delta h\right) - {1\over2}\left(n | n\right)\right]
\,{\cal D}n\;,
\label{appA:prob1}
\end{equation}
where the ``one-detector'' inner product is
\begin{equation}
(a|b) = 4\,{\rm Re}\int_0^\infty
{{\tilde a}(f)^*{\tilde b}(f)\over S_h(f)}\,df\;.
\label{appA:innerprod1}
\end{equation}

Following Finn {\cite{finn92}}, we evaluate the functional integral
(\ref{appA:prob1}) by treating the datastream in the limit of discrete
sampling.  Take the total span of data to be of finite duration $T$.
Let the sampling time be $\Delta t$, and define
\begin{eqnarray}
t_i &=& i \Delta t, \qquad i = 0,\ldots,{\cal N}\;,
\nonumber\\
a_i &=& a(t_i)\;,
\nonumber\\
C_{n,ij} &=& C_n(t_i - t_j)\;.
\label{appA:defs1}
\end{eqnarray}
The function $C_n(t)$ is the noise autocorrelation,
\begin{equation}
C_n(\tau) = {\rm E}\{n(t)n(t+\tau)\}_n\;,
\end{equation}
related to the spectral density by
\begin{equation}
S_h(f) = 2\int_{-\infty}^\infty e^{2\pi i f \tau}C_n(\tau)\,d\tau\;.
\end{equation}

Let ${\bf a}$ be the vector whose components are $a_i$, and let ${\bf
C}_n$ be the matrix whose components are $C_{n,ij}$.  Note that ${\bf
C}_n$ is a real, symmetric matrix.  The inner product
(\ref{appA:innerprod1}) becomes [{\it cf.} \cite{finn92}, Eq.\ (2.20)]
\begin{equation}
(a|b) = \lim_{{\mbox{$\stackrel{\textstyle\Delta t\to0}{T\to\infty}$}}}
	{\bf a}\cdot{\bf C}_n^{-1}\cdot{\bf b}\;.
\label{appA:innerprod2}
\end{equation}
The probability distribution (\ref{appA:prob1}) can therefore be written
\begin{equation}
P_{\rm post}(\lambda|\hat\lambda) =
\lim_{{\mbox{$\stackrel{\textstyle \Delta t\to0}{T\to\infty}$}}}
p_0(\lambda)\int \exp\left[-{\bf n}\cdot{\bf C}_n^{-1}\cdot{\bf n}
-{\bf n}\cdot{\bf C}_n^{-1}\cdot\delta{\bf h}
-{1\over2}\delta{\bf h}\cdot{\bf C}_n^{-1}\cdot\delta{\bf h}\right]\,
d{\bf n}\;.
\label{appA:prob2}
\end{equation}
This integral is of a form commonly encountered in path integral
quantization, and evaluates to
\begin{eqnarray}
P_{\rm post}(\lambda | \hat\lambda) &=&
\lim_{{\mbox{$\stackrel{\textstyle \Delta t\to0}{T\to\infty}$}}}
p_0(\lambda)\exp\left[-{1\over4}\delta{\bf h}\cdot
{\bf C}_n^{-1}\cdot\delta{\bf h}\right]
\nonumber\\
&=& p_0(\lambda)\exp\left[-{1\over4}(\delta h | \delta h)\right]\;.
\end{eqnarray}
[We have absorbed a constant proportional to $1/\sqrt{\det{{\bf
C_n}}}$ into the prior probability $p_0(\lambda)$.]

Now consider the entire detector network:
\begin{equation}
P_{\rm post}(\lambda|\hat\lambda) =
{\cal K}p_0(\lambda)
\int\exp\left[-{1\over2}\left({\vec n}|{\vec n}\right)\right]
\exp\left[-{1\over2}\left(\delta{\vec h}|\delta{\vec h}\right)
-\left({\vec n}|\delta{\vec h}\right)
-{1\over2}\left({\vec n}|{\vec n}\right)\right]\,{\cal D}{\vec n}\;.
\label{appA:prob3}
\end{equation}
Recall that the ``network'' inner product is
\begin{equation}
({\vec a} | {\vec b}) \equiv 4\,{\rm Re} \int_0^\infty df\;
{\tilde a}_i(f)^*\left[{\bf S}_h(f)^{-1}\right]^{ij}{\tilde b}_j(f)\;.
\label{appA:innerprod3}
\end{equation}
Because the matrix ${\bf S}_h(f)^{-1}$ is real and symmetric, its
eigenvectors are orthonormal.  If the matrix of eigenvectors is ${\bf
A}$ then ${\bf A}^{-1}={\bf A}^T$, and the matrix
\begin{equation}
{\bf Z}_h(f)^{-1} = {\bf A}\cdot{\bf S}_h(f)^{-1}\cdot{\bf A}^T
\label{appA:diagonal}
\end{equation}
is diagonal.  Define further
\begin{eqnarray}
{\tilde a}^\prime_j(f)^* &=& {\tilde a}_i(f)^* A^i_j\;,
\nonumber\\
{\tilde b}^\prime_j(f) &=& A^i_j {\tilde b}_i(f)\;.
\label{appA:defs2}
\end{eqnarray}
Using Eqs.\ (\ref{appA:diagonal}) and (\ref{appA:defs2}), the inner
product becomes
\begin{eqnarray}
({\vec a} | {\vec b})
&=& 4\,{\rm Re} \int_0^\infty
{\tilde a}_n(f)^* A^n_k A^k_i
\left[{\bf S}_h(f)^{-1}\right]^{ij}
A^l_j A^m_l{\tilde b}_m(f)\,df
\nonumber\\
&=& 4\,{\rm Re} \int_0^\infty
{\tilde a}^\prime_k(f)^*
\left[{\bf Z}_h^{-1}(f)\right]^{kl}
{\tilde b}^\prime_l(f)\,df
\nonumber\\
&=& \sum_k 4\,{\rm Re} \int_0^\infty
{\tilde a}^\prime_k(f)^*
\left[{\bf Z}_h^{-1}(f)\right]^{kk}
{\tilde b}^\prime_k(f)\,df
\nonumber\\
&\equiv& \sum_k\langle a_k^\prime | b_k^\prime \rangle_k\;.
\label{appA:innerprod4}
\end{eqnarray}
The third equality follows from the fact that ${\bf Z}_h^{-1}(f)$ is
diagonal.  The inner product $\langle a_k^\prime |b_k^\prime\rangle_k$
is identical to the inner product (\ref{appA:innerprod1}) except that
$[{\bf Z}_h^{-1}]^{kk}$ is used in place of $1/S_h(f)$.  Using
(\ref{appA:innerprod4}), Eq.\ (\ref{appA:prob3}) becomes
\begin{eqnarray}
P_{\rm post}(\lambda|\hat\lambda)
&=& {\cal K}p_0(\lambda)\int\exp\left[-\sum_k\left({1\over2}
\langle\delta h^\prime_k | \delta h^\prime_k\rangle_k +
\langle n^\prime_k | \delta h^\prime_k\rangle_k +
\langle n^\prime_k | n^\prime_k\rangle_k\right)\right]\,
{\cal D}{\vec n}^\prime
\nonumber\\
&=& {\cal K}p_0(\lambda)\prod_k\int\exp\left[-{1\over2}
\langle\delta h^\prime_k | \delta h^\prime_k\rangle_k -
\langle n^\prime_k | \delta h^\prime_k\rangle_k -
\langle n^\prime_k | n^\prime_k\rangle_k\right]\,
{\cal D}n^\prime_k\;.
\label{appA:prob4}
\end{eqnarray}
Since the matrix ${\bf A}$ is unitary, the functional differential
element ${\cal D}{\vec n}^\prime = {\cal D}{\vec n}$.  The integral on
the last line of (\ref{appA:prob4}) is identical to the integral in
Eq.\ (\ref{appA:prob1}), except for the slightly different inner
product.  The result is therefore
\begin{eqnarray}
P_{\rm post}(\lambda | \hat\lambda)
&=& {\cal K}p_0(\lambda)\prod_k
\exp\left[-{1\over4}\langle\delta h^\prime_k|
\delta h^\prime_k\rangle_k\right]
\nonumber\\
&=& {\cal K}p_0(\lambda)\exp\left[-{1\over4}\sum_k
\langle\delta h^\prime_k | \delta h^\prime_k\rangle_k\right]
\nonumber\\
&=& {\cal K}p_0(\lambda)\exp\left[-{1\over4}
(\delta{\vec h} | \delta{\vec h})\right]\;.
\label{appA:prob5}
\end{eqnarray}
This is the result claimed in Eq.\ (\ref{eq:postprob}).

\begin{table}
\begin{tabular}{lclclcl}
$N_N$ && $\rho_{\rm insp} = 10$ && $\rho_{\rm insp} = 30$
&& $\rho_{\rm insp} = 90$ \\
\colrule
1 && $f_1 = 630$ Hz && $f_1 = 710$ Hz && $f_1 = 1380$ Hz \\
\colrule
2 && $f_1 = 620$ Hz, $f_2 = 670$ Hz && $f_1 = 620$ Hz, $f_2 = 780$ Hz
  && $f_1 = 760$ Hz, $f_2 = 1390$ Hz \\
\colrule
3 && $f_1 = 600$ Hz, $f_2 = 660$ Hz,&& $f_1 = 590$ Hz, $f_2 = 690$ Hz,
  && $f_1 = 740$ Hz, $f_2 = 1340$ Hz,\\ 
  && $f_3 = 700$ Hz&& $f_3 = 820$ Hz&& $f_3 = 1420$ Hz\\
\colrule
4 && $f_1 = 590$ Hz, $f_2 = 650$ Hz,&& $f_1 = 580$ Hz, $f_2 = 670$ Hz,
  && $f_1 = 650$ Hz, $f_2 = 770$ Hz,\\
  && $f_3 = 700$ Hz, $f_4 = 730$ Hz && $f_3 = 790$ Hz, $f_4 = 1390$ Hz,
  && $f_3 = 1350$ Hz, $f_4 = 1430$ Hz\\
\colrule
5 && $f_1 = 580$ Hz, $f_2 = 630$ Hz,&& $f_1 = 580$ Hz, $f_2 = 660$ Hz,
  && $f_1 = 660$ Hz, $f_2 = 760$ Hz,\\
  && $f_3 = 680$ Hz, $f_4 = 710$ Hz, &&$f_3 = 780$ Hz, $f_4 = 1350$ Hz,
  && $f_3 = 1320$ Hz, $f_4 = 1400$ Hz,\\
  && $f_5 = 740$ Hz&& $f_5 = 1400$ Hz &&$f_5 = 1450$ Hz\\
\colrule
\end{tabular}
\caption
{Optimal detector networks in the toy description for uniform prior
probability distribution.  This table shows the central frequencies
for networks with one broad-band and $N_N$ narrow-band detectors, for
several values of inspiral SNR.
}
\label{tab:network}
\end{table}

\begin{figure}
\epsfig{file = 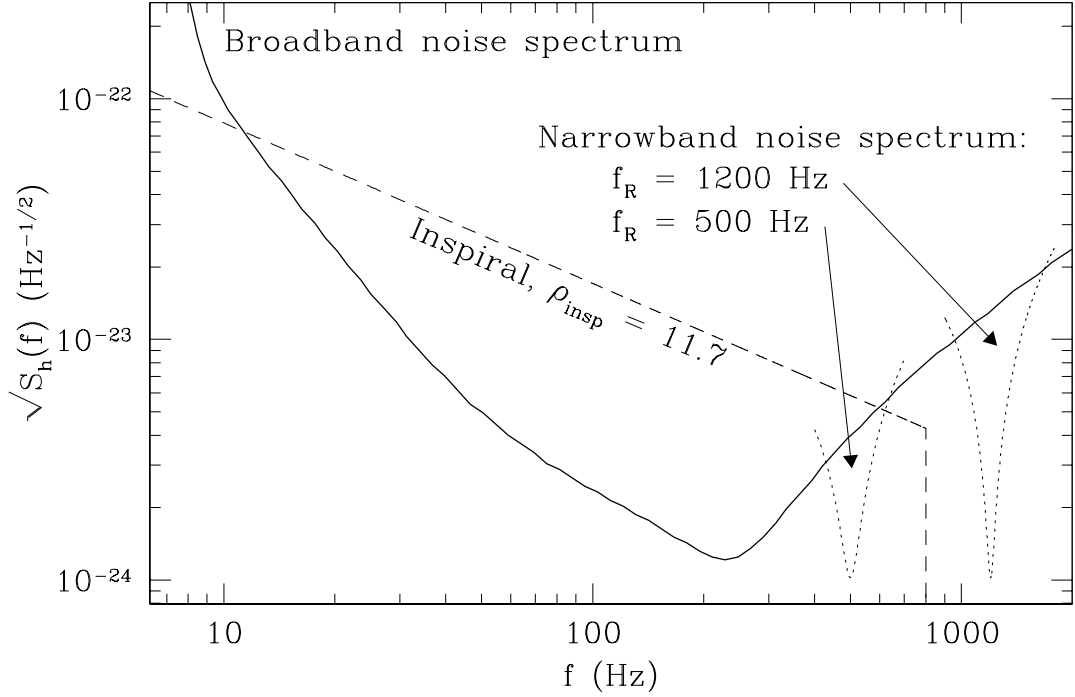, width = 15cm}
\vskip -2.7cm
\caption{Broad-band noise used in this analysis (second generation
LIGO sensitivity, optimized to detect binary neutron star inspiral),
with two examples of narrow-band interferometer noise.  These examples
are for resonant frequencies of $500$ Hz and $1200$ Hz; each assumes a
bandwidth of $50$ Hz.  The dashed line is a representation of the wave
strain for a signal whose inspiral SNR is 11.7 (corresponding to an
optimally oriented source at 500 Mpc, or an ``average'' source at 200
Mpc).
}
\label{fig:noise}
\end{figure}

\begin{figure}
\epsfig{file = 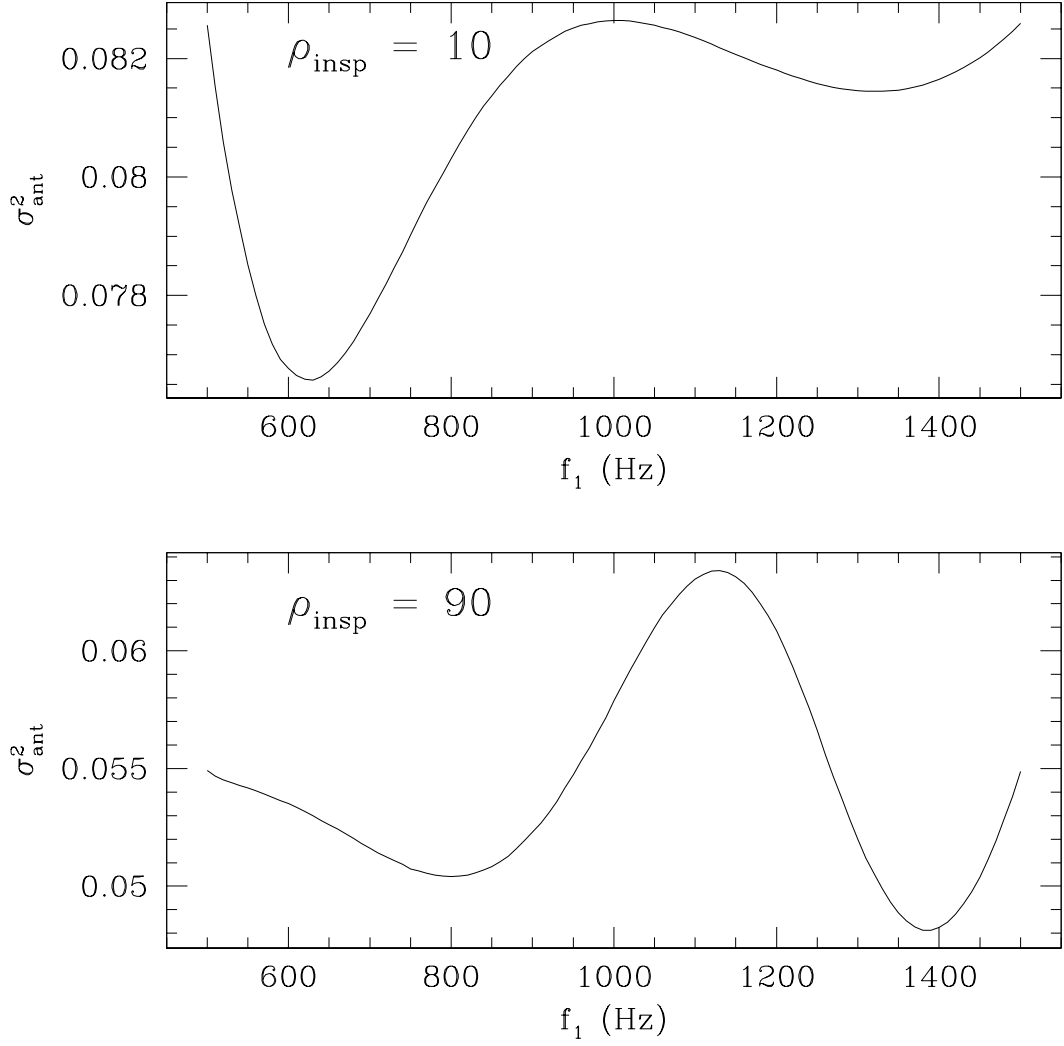, width = 15cm}
\caption{The anticipated variance for measurements with a single
narrow-band detector at inspiral SNR $\rho_{\rm insp} = 10$ and
$\rho_{\rm insp} = 90$.  At relatively low SNR, the tuning algorithm
places the narrow-band detector at $f\simeq 620\,{\rm Hz}$ to avoid
false positives, since it knows the signal will be extremely weak at
high frequencies.  At high SNR, the algorithm places the detector at
$f\simeq 1400\,{\rm Hz}$ --- the risk of a false positive is much
lower, so the algorithm chooses the narrow-band detector to counter
the rapidly growing broad-band noise.
}
\label{fig:var}
\end{figure}

\begin{figure}
\epsfig{file = 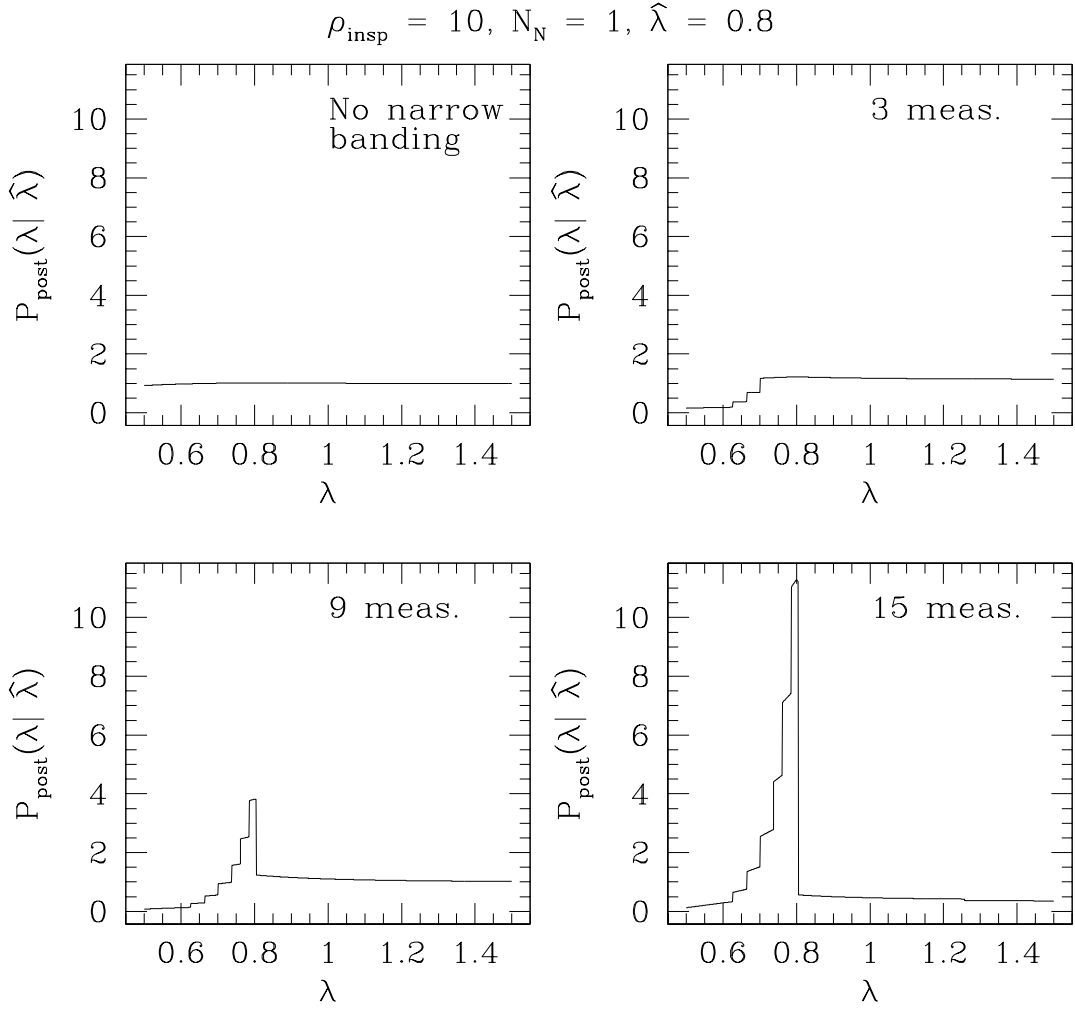, width = 15cm}
\caption{The evolution of the posterior probability distribution,
measuring coalescences with $\rho_{\rm insp} = 10$, $\hat\lambda =
0.8$, and a single narrow-band detector.  After each measurement, we
update the prior distribution, $p_0^i(\lambda) = P_{\rm post}^{i -
1}(\lambda |\hat\lambda)$, and then retune the network (minimizing
$\sigma_{\rm ant}^2$).  In this way, multiple measurements drive the
network to ``zoom in'' on a distribution that is peaked about the true
value $\hat\lambda$.  In the case shown here, the distribution starts
to become peaked after about 9 measurements, and has a pronounced peak
after 15 measurements.
}
\label{fig:snr10_nb1_lh0.8}
\end{figure}

\begin{figure}
\epsfig{file = 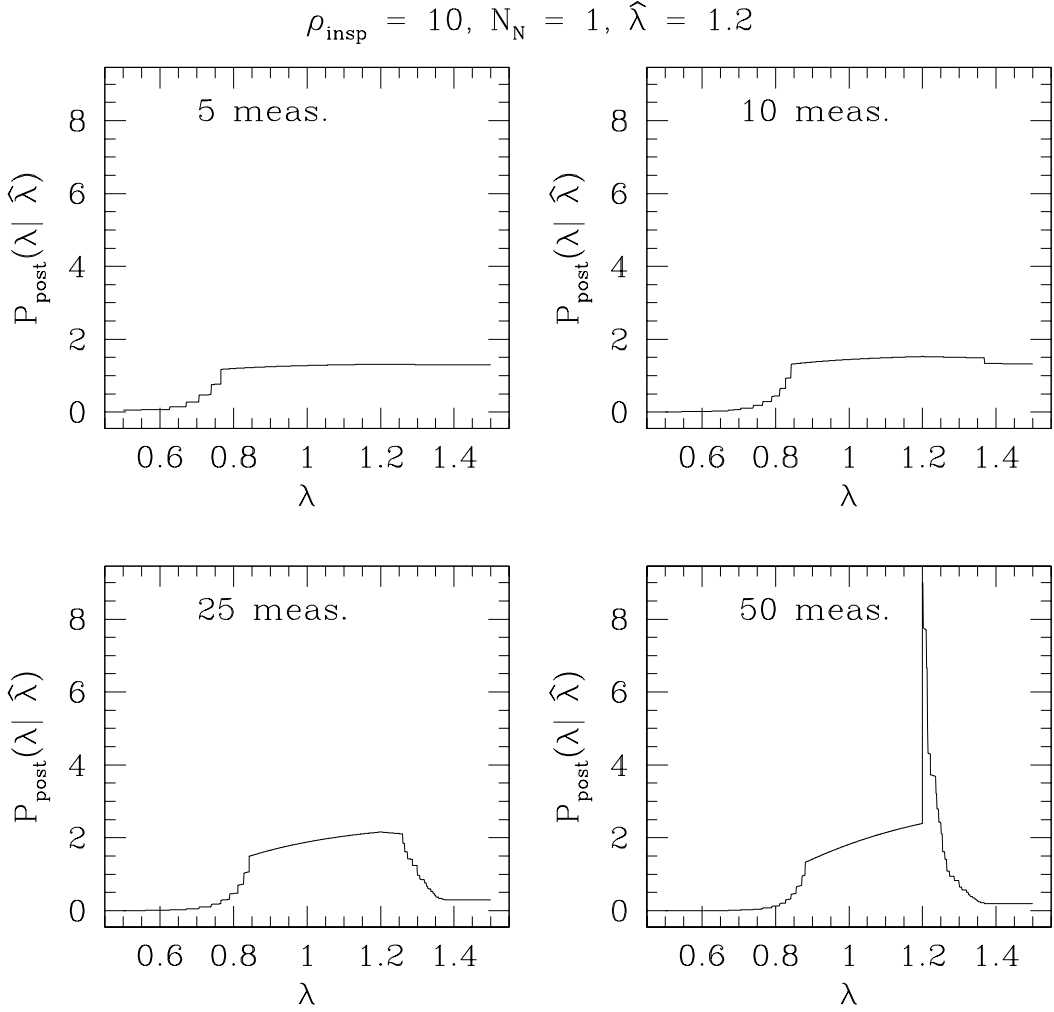, width = 15cm}
\caption{The same as Fig.\ {\ref{fig:snr10_nb1_lh0.8}}, but with
$\hat\lambda = 1.2$.  The evolution of the posterior probability is
qualitatively similar to the evolution when $\hat\lambda = 0.8$, but
is much slower.  At this signal strength, the tuning algorithm tends
not to sample the high-frequency end of its parameter space, in order
to reduce the probability of false positive measurements.  Because the
merger occurs at high frequency in this case, the algorithm takes
quite a while to find the merger power.
}
\label{fig:snr10_nb1_lh1.2}
\end{figure}

\begin{figure}
\epsfig{file = 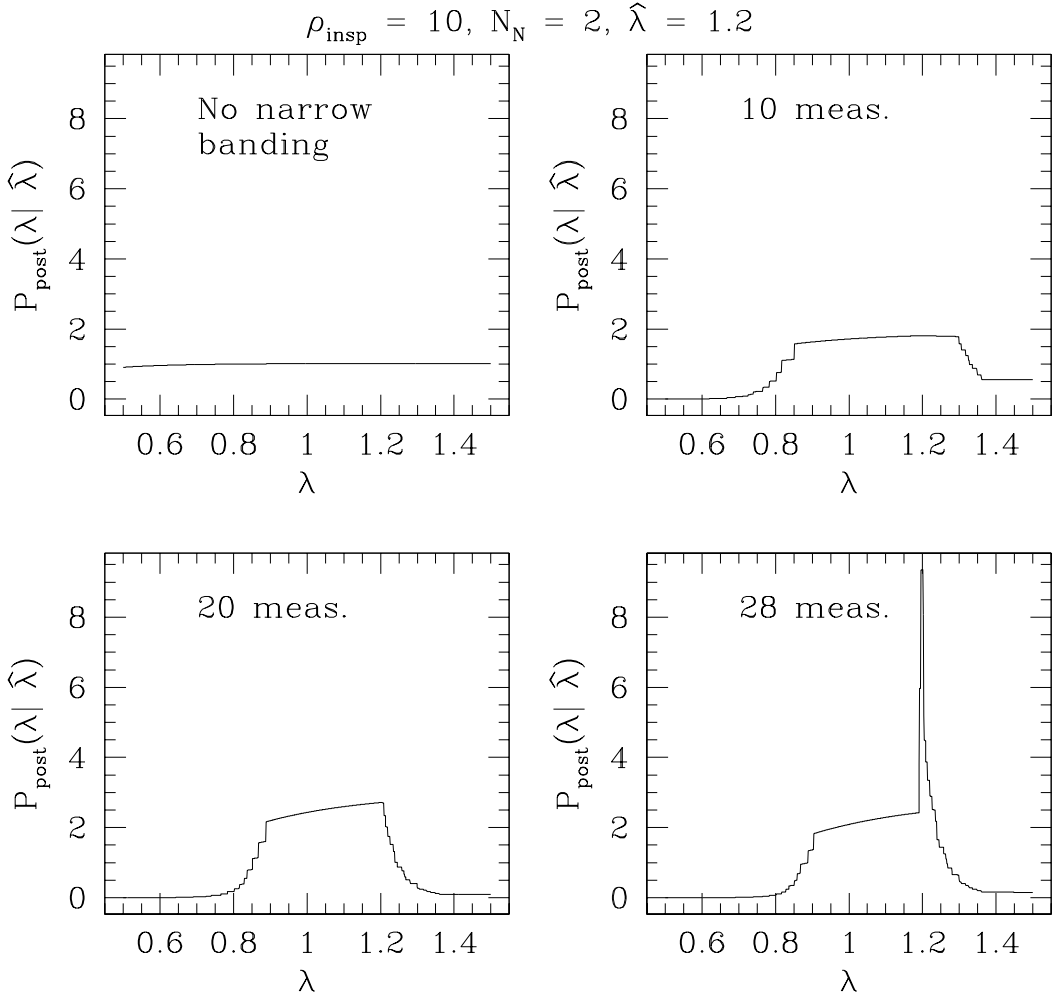, width = 15cm}
\caption{The same as Fig.\ {\ref{fig:snr10_nb1_lh1.2}}, but using
two narrow-band detectors.  Convergence onto a distribution peaked
near $\hat\lambda$ is still slow, but not as bad as when a single
detector is used.
}
\label{fig:snr10_nb2_lh1.2}
\end{figure}

\begin{figure}
\epsfig{file = 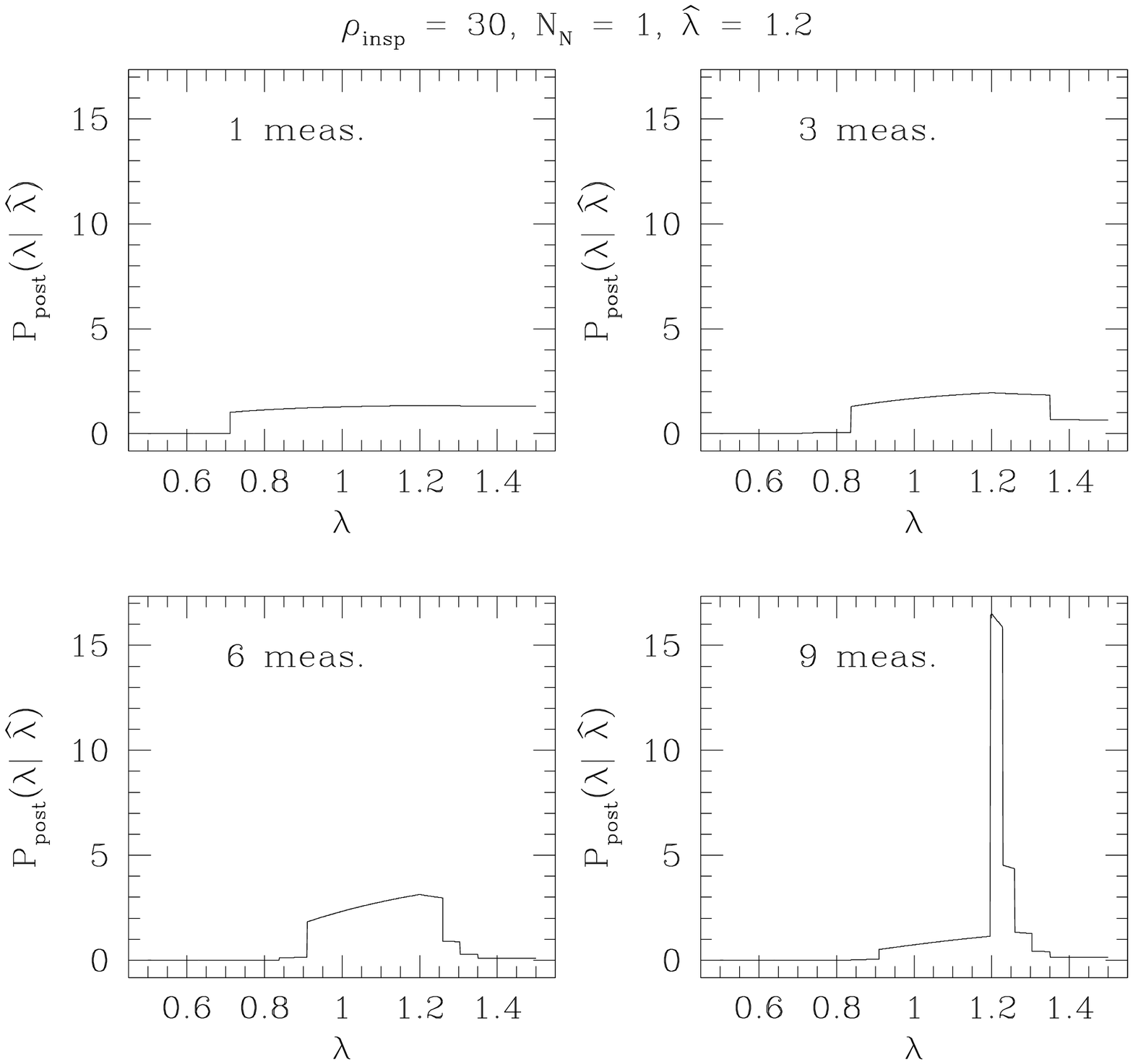, width = 15cm}
\caption{The same as Fig.\ {\ref{fig:snr10_nb1_lh1.2}}, but at higher
SNR: $\rho_{\rm insp} = 30$.  SNR has a marked effect on the rate at
which the distributions converge onto $\hat\lambda$.  The algorithm
does not need to worry about false positives so much in this case, and
so it is quite happy to assign the single detector to high frequencies
when necessary.  As a consequence, the distribution is quite peaked
after only 9 measurements --- in marked contrast to the 50
measurements needed when $\rho_{\rm insp} = 10$.  }
\label{fig:snr30_nb1_lh1.2}
\end{figure}


\begin{thebibliography}{99}

\bibitem{clark_eardley} J.\ P.\ A.\ Clark and D.\ M.\ Eardley,
  Astrophys.\ J.\ {\bf 215}, 311 (1977).

\bibitem{300yrs} K.\ S.\ Thorne, in {\it Three Hundred Years of
  Gravitation}, edited by S.\ Hawking and W.\ Israel (Cambridge
  University Press, Cambridge, England, 1987), p.\ 330.

\bibitem{3minutes} C.\ Cutler et al., Phys.\ Rev.\ Lett.\ {\bf 70},
  2984 (1993).

\bibitem{vallis} M.\ Vallisneri, Phys.\ Rev.\ Lett.\ {\bf 84}, 3519
  (2000).

\bibitem{ligo} See {\tt http://www.ligo.caltech.edu} and links therein
  for a discussion of the current state of LIGO.

\bibitem{klt} D.\ Kennefick, D.\ Laurence, and K.\ S.\ Thorne, in {\it
  Proceedings of the 7th Marcel Grossman Meeting}, edited by R.\ T.\
  Jantzen and G.\ Mac Keiser (World Scientific, Singapore, 1997), p.\
  1090.

\bibitem{lee_eos} L.\ Lindblom, Astrophys.\ J.\ {\bf 398}, 569 (1992).

\bibitem{alford} M.\ Alford, Lect.\ Notes Phys.\ {\bf 583}, 81 (2002).

\bibitem{arrw} M.\ G.\ Alford, K.\ Rajagopal, S.\ Reddy, and
  F.\ Wilczek, Phys.\ Rev.\ D {\bf 64}, 073017 (2001).

\bibitem{zcm} X.\ Zhuge, J.\ M.\ Centrella, and S.\ L.\ W.\ McMillan,
  Phys.\ Rev.\ D {\bf 50}, 6247 (1994); {\bf 54}, 7261 (1996).

\bibitem{rs} F.\ A.\ Rasio and S.\ L.\ Shapiro, Class.\ Quantum Grav.\
  {\bf 16}, R1 (1999).

\bibitem{fgrt} J.\ A.\ Faber, P.\ Grandcl\'ement, F.\ A.\ Rasio, and
  K.\ Taniguchi, Phys.\ Rev.\ Lett., submitted; also astro-ph/0204397.

\bibitem{blair} D.\ G.\ Blair, Class.\ Quantum Grav.\ {\bf 18}, 4087
  (2001).

\bibitem{cerdonio} M.\ Cerdonio, Class.\ Quantum Grav.\ {\bf 18}, 4101
	(2001).

\bibitem{sr} B.\ J.\ Meers, Phys.\ Rev.\ D {\bf 38}, 2317 (1988).

\bibitem{rse} J.\ Mizuno, K.\ A.\ Strain, P.\ G.\ Nelson, J.\ M.\
  Chen, R.\ Schilling, A.\ R\"udiger, W.\ Winkler, and K.\ Danzmann,
  Phys.\ Lett.\ A {\bf 175}, 273 (1993).

\bibitem{finn92} L.\ S.\ Finn, Phys.\ Rev.\ D {\bf 46}, 5236 (1992).

\bibitem{finnchernoff} L.\ S.\ Finn and D.\ F.\ Chernoff, Phys.\ Rev.\
  D {\bf 47}, 2198 (1993).

\bibitem{cf} C.\ Cutler and E.\ E.\ Flanagan, Phys.\ Rev.\ D {\bf 49},
  2658 (1994).

\bibitem{whitepaper} E.\ Gustafson, D.\ Shoemaker, K.\ Strain, and R.\
  Weiss, {\it LSC White Paper on Detector Research and Development},
  LIGO Project Document T990080-00-D (1999).

\bibitem{hhs} G.\ M.\ Harry, J.\ L.\ Houser, and K.\ A.\ Strain,
  Phys.\ Rev.\ D {\bf 65}, 082001 (2002).

\bibitem{pw} E.\ Poisson and C.\ M.\ Will, Phys.\ Rev.\ D {\bf 52},
  848 (1995).

\bibitem{bbhII} E.\ E.\ Flanagan and S.\ A.\ Hughes, Phys.\ Rev.\ D
  {\bf 57}, 4566 (1998).

\bibitem{bkb2001} K.\ Belczynski, V.\ Kalogera, and T.\ Bulik,
  Astrophys.\ J.\ {\bf 572}, 407 (2001).

\bibitem{kkl2002} C.\ Kim, V.\ Kalogera, and D.\ R.\ Lorimer,
  Astrophys.\ J., submitted (astro-ph/0207408).

\bibitem{bench} The program {\sc bench} can be obtained from the URL
  {\tt http://gravity.phys.psu.edu/Bench}

\bibitem{klm} A.\ Kr\'olak, J.\ A.\ Lobo, and B.\ Meers, Phys.\ Rev.\
  D {\bf 43}, 2470 (1991).

\bibitem{krolak} A.\ Kr\'olak, Acta Cosmologica {\bf 22-1}, 1 (1996).

\bibitem{ds_private} D.\ Shoemaker, private communication.

\bibitem{bc2001} A.\ Buonanno and  Y.\ Chen, Class.\ Quantum Grav.\
  {\bf 18}, L95 (2001).

\bibitem{recipes} W.\ H.\ Press, S.\ A.\ Teukolsky, W.\ T.\
  Vetterling, and B.\ P.\ Flannery, {\it Numerical Recipes} (Cambridge
  University Press, Cambridge, 1992).

\bibitem{phd} S.\ A.\ Hughes, unpublished Ph.\ D.\ thesis, Caltech,
  1998.

\end{thebibliography}
\end{document}